\documentclass[lettersize,journal]{IEEEtran}
\usepackage{amsmath,amsfonts}
\usepackage{algorithmic}
\usepackage{array}
\usepackage[caption=false,font=normalsize,labelfont=sf,textfont=sf]{subfig}
\usepackage{textcomp}
\usepackage{stfloats}
\usepackage{url}
\usepackage{verbatim}
\usepackage{graphicx}
\usepackage{cite}
\usepackage{gensymb}
\usepackage{hyperref}
\usepackage{orcidlink}
\def\BibTeX{{\rm B\kern-.05em{\sc i\kern-.025em b}\kern-.08em
    T\kern-.1667em\lower.7ex\hbox{E}\kern-.125emX}}
\usepackage{balance}
\begin{document}
\title{The Simons Observatory: Studies of Phase Drift in RF Transmission Lines from the First Large-Scale Deployment of Microwave Frequency Multiplexing for Cosmology}

\author{Thomas P. Satterthwaite\orcidlink{0000-0002-6452-4220}, Zeeshan Ahmed\orcidlink{0000-0002-9957-448X}, Cody J. Duell\orcidlink{0000-0002-6318-1924}, Shawn W. Henderson\orcidlink{0000-0001-7878-4229}, Tristan Pinsonneault-Marotte\orcidlink{0000-0002-9516-3245}, Max Silva-Feaver\orcidlink{0000-0001-7480-4341}, Yuhan Wang\orcidlink{0000-0002-8710-0914}
\thanks{T. P. Satterthwaite is with the Department of Physics, Stanford University, USA and the Kavli Institute for Particle Astrophysics \& Cosmology, Stanford, USA.}
\thanks{Z. Ahmed is with the Kavli Institute for Particle Astrophysics \& Cosmology, Stanford, USA and SLAC National Accelerator Laboratory, Menlo Park, USA.}
\thanks{C. J. Duell is with the Department of Physics, Cornell University, USA.}
\thanks{S. W. Henderson is with the Kavli Institute for Particle Astrophysics \& Cosmology, Stanford, USA and SLAC National Accelerator Laboratory, Menlo Park, USA.}
\thanks{T. Pinsonneault-Marotte is with the Kavli Institute for Particle Astrophysics \& Cosmology, Stanford, USA and SLAC National Accelerator Laboratory, Menlo Park, USA.}
\thanks{M. Silva-Feaver is with the Wright Laboratory, Department of Physics, Yale University, USA.}
\thanks{Y. Wang is with the Department of Physics, Cornell University, USA.}}

\maketitle

\begin{abstract}
Fulfilling the science goals of the Simons Observatory, a state-of-the-art cosmic microwave background (CMB) experiment, has required deploying tens of thousands of superconducting bolometers. Reading out data from the observatory's more than 67,000 transition-edge sensor (TES) detectors while maintaining cryogenic conditions requires an effective multiplexing scheme. The SLAC microresonator radio frequency (SMuRF) electronics have been developed to provide the warm electronics for a high-density microwave frequency multiplexing readout system, and this system has been shown to achieve multiplexing factors on the order of 1,000. SMuRF has recently been deployed to the Simons Observatory, which is located at 5,200\;m on Cerro Toco in Chile’s Atacama Desert. As the SMuRF system is exposed to the desert’s diurnal temperature swings, resulting phase drift in RF transmission lines may introduce a systematic signal contamination. We present studies of phase drift in the room-temperature RF lines of the Simons Observatory’s 6\;m large-aperture telescope, which hosts the largest deployment to date of TES microwave frequency multiplexing to a single telescope. We show that these phase drifts occur on time scales which are significantly longer than sky scanning, and that their contribution to on-sky in-transition detector noise is within the readout noise budget.
\end{abstract}


\section{Introduction}
\label{sec:intro}

\IEEEPARstart{T}{he} cosmic microwave background (CMB)---relic light from the era of recombination---is a powerful probe of fundamental physics. Measurements of this light provide effective tests of the growth of large-scale structure, constraints on cosmological parameters, and bounds on the sum of the masses of neutrinos, while its polarization signature may tell of the energy scale of cosmic inflation\cite{CMBScience-CMBS4_2016}. Observing the CMB with greater precision from ground-based observatories requires deploying large numbers of photon-noise-limited bolometers. To this end, the Simons Observatory has recently deployed more than 67,000 superconducting transition-edge sensor (TES) detectors to four telescopes located on Cerro Toco at 5,200\;m in Chile's Atacama Desert. Its three 0.5\;m refracting small-aperture telescopes (SATs) each contain more than 12,000 of these detectors with the goal of mapping 10\% of the sky with $2\;\mu\text{K-arcmin}$ map depth in CMB bands, while its 6\;m crossed-Dragone large-aperture telescope (LAT) contains nearly 31,000 detectors with the goal of mapping 40\% of the sky with $6\;\mu\text{K-arcmin}$ map depth in CMB bands\cite{SO-Ade_2019,SATs-Galitzki_2024,LAT-Xu_2021,LATR-Zhu_2021,LATRDark-Bhandarkar_2025}. A forthcoming upgrade will deploy nearly 32,000 additional detectors to the LAT as part of the Advanced Simons Observatory upgrade, and additional SATs with tens of thousands of TES and microwave kinetic inductance detectors are planned\cite{ASO-Simons_2025,Enhanced-Lee_2019}.

As these tens of thousands of detectors operate at 100\;mK, an improved multiplexing scheme is necessary to read out their signals. This is accomplished using microwave frequency multiplexing, and the deployment of this technology to the Simons Observatory's telescopes has been the largest deployment of TES microwave frequency multiplexing for cosmology to date\cite{SQUIDMUX-Irwin_2004,uMUX-Mates_2011,Satterthwaite-SMuRFDeployment}. Using SLAC microresonator radio frequency (SMuRF) warm electronics and purpose-built cryogenic universal focal-plane modules (UFMs), the Simons Observatory has achieved a multiplexing factor on the order of 1,000 per pair of radio frequency (RF) transmission lines\cite{SMuRF-Yu_2023,UFM-McCarrick_2021}. In this system, RF probe tones generated by SMuRF are carried by transmission lines which enter the instruments' cryostats and interrogate microwave resonators coupled to TESs on their focal planes. The room-temperature transmission lines, which are external to the cryostat, however, are subject to changes in ambient temperature, which may introduce a systematic phase drift as their lengths and dielectric constants are altered\cite{ToneTracking-SilvaFeaver_2022}. If this effect is not characterized and mitigated, it may lead to a contamination by mimicking TES detector signal.

The SMuRF electronics for the SATs are installed inside water-cooled enclosures which provide isolation from the ambient environment\cite{SATs-Galitzki_2024}. For the LAT, however, the electronics are mounted directly to the LAT receiver (LATR) which is installed to the telescope's receiver cabin\cite{LATR-Zhu_2021}. This cabin of the crossed-Dragone telescope is kept isothermal with the telescope's mirrors for optical stability, meaning that the warm electronics and their transmission lines are exposed to diurnal temperature swings\cite{CCAT-Parshley_2018}. While previous work has elucidated the importance of using phase-stable transmission lines and tone-tracking firmware to mitigate the coupling of these temperature swings to signal, the efficacy of these methods has not yet been demonstrated in the field\cite{Cables-Wang_2024}\cite{ToneTracking-SilvaFeaver_2022}.

This paper is organized as follows: Section \ref{sec:smurf} describes the SMuRF electronics and their installation to the LAT, and Section \ref{sec:fixed-tones} presents the off-resonance tones used to measure the scale of diurnal phase drifts during observations and the projected impact of this effect on data quality.

\section{SMuRF Electronics for the LAT}
\label{sec:smurf}
\noindent

\begin{figure}[!t]
\centering
\includegraphics[width=2.5in]{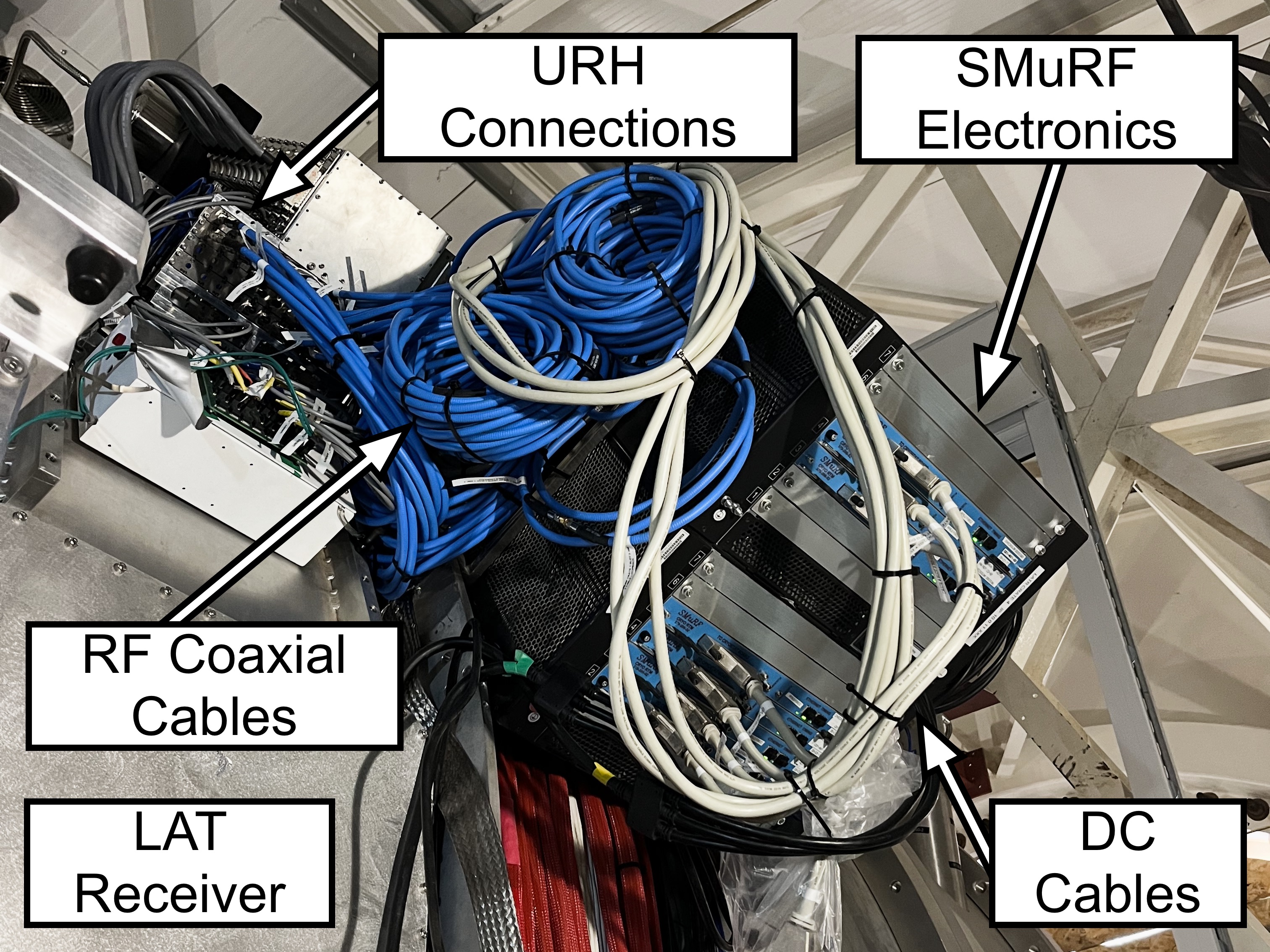}
\caption{Two ATCA crates containing a total of seven SMuRF systems installed to the LATR. RF coaxial cables and DC cables, which transmit signals into the cryostat via the URH, are shown.}
\label{fig:smurf}
\end{figure}

SMuRF electronics provide the warm system that enables microwave frequency multiplexing at the Simons Observatory. Each SMuRF system consists of a carrier board which drives two Advanced Mezzanine Cards (AMCs) that produce and demodulate 4-6\;GHz probe tones to interrogate the cryogenic microwave resonators, and a Rear Transition Module (RTM) that generates TES bias line voltages, warm amplifier biases, and flux ramp signals for linearizing the TES response, which enter the cryostat via a cryostat card. The carrier board, AMC, and RTM systems are installed to Advanced Telecommunications Computing Architecture (ATCA) crates along with network switches that facilitate communication with data acquisition servers and timing systems that interface with an observatory-wide 122.88\;MHz clock. The cryostat cards are installed to a separate aluminum enclosure which is mounted adjacent to the ATCA crates. For a more detailed discussion of the SMuRF electronics, refer to \cite{SMuRF-Yu_2023}.

Universal focal-plane modules (UFMs) contain the cryogenic multiplexing circuit which consists of feedhorn-coupled TES detectors that are inductively coupled to superconducting quantum interference devices (SQUIDs), which are in turn coupled to microwave frequency resonators. For a more detailed discussion of the UFMs and their performance, refer to \cite{UFM-McCarrick_2021,220280UFM-Healy_2022,90150UFM-Dutcher_2023}. Each SMuRF system interfaces with a single UFM and its 1,720 optically-coupled TES detectors, with each AMC reading out half of the UFM. The system therefore reaches a multiplexing factor of 860 for optically-coupled detectors.

For the Simons Observatory's LAT, the ATCA crates and cryocard enclosures are mounted directly onto the LATR so that the SMuRF systems and their cables move with the receiver as it rotates in boresight. A total of 18 SMuRF systems deployed to four ATCA crates have thus far been installed to the LATR in order to read out its 18 UFMs. Four 2\;m True Blue 205 Ruggedized SMA coaxial cables\footnote{Part number 90-010-2MTR.}---chosen for their relative phase stability---and one set of DC cables per SMuRF system carry signals into and out of the cryostat via universal readout harnesses (URHs) which provide the interface into the cryogenic section of the receiver\cite{URH-Moore_2022}. Figure \ref{fig:smurf} shows two ATCA crates installed to the LATR, with their cables and the URH connections labeled.

In order to maintain optical stability between the LATR and the telescope's mirrors, the receiver cabin is kept isothermal with the rest of the telescope's structure. This implies that the cabin and its SMuRF systems are exposed to the diurnal temperature swings on Cerro Toco. While the electronics use an active air cooling system from the ATCA crates to dissipate heat produced by their field-programmable gate arrays, the temperatures of the coaxial cables are not regulated. Changes in the cable lengths and dielectric properties during these diurnal swings may therefore introduce a systematic signal contamination to measurements made by the SMuRF system via a phase drift.

\section{Phase Stability of Off-Resonance Tones}
\label{sec:fixed-tones}
\noindent

In order to probe the magnitude of these phase drifts, we analyze off-resonance tones produced by the SMuRF systems. Each AMC produces tones in a 2\;GHz range, split into four 500\;MHz-wide bands whose constituent tones are generated by a dedicated DAC. A closed-loop tone-tracking algorithm adjusts these tones as the cryogenic microwave frequency resonators respond to a flux ramp signal and incident TES loading. One ``on-resonance" tone is produced and tone-tracked per detector.

\begin{figure}[!t]
\centering
\includegraphics[width=3.25in]{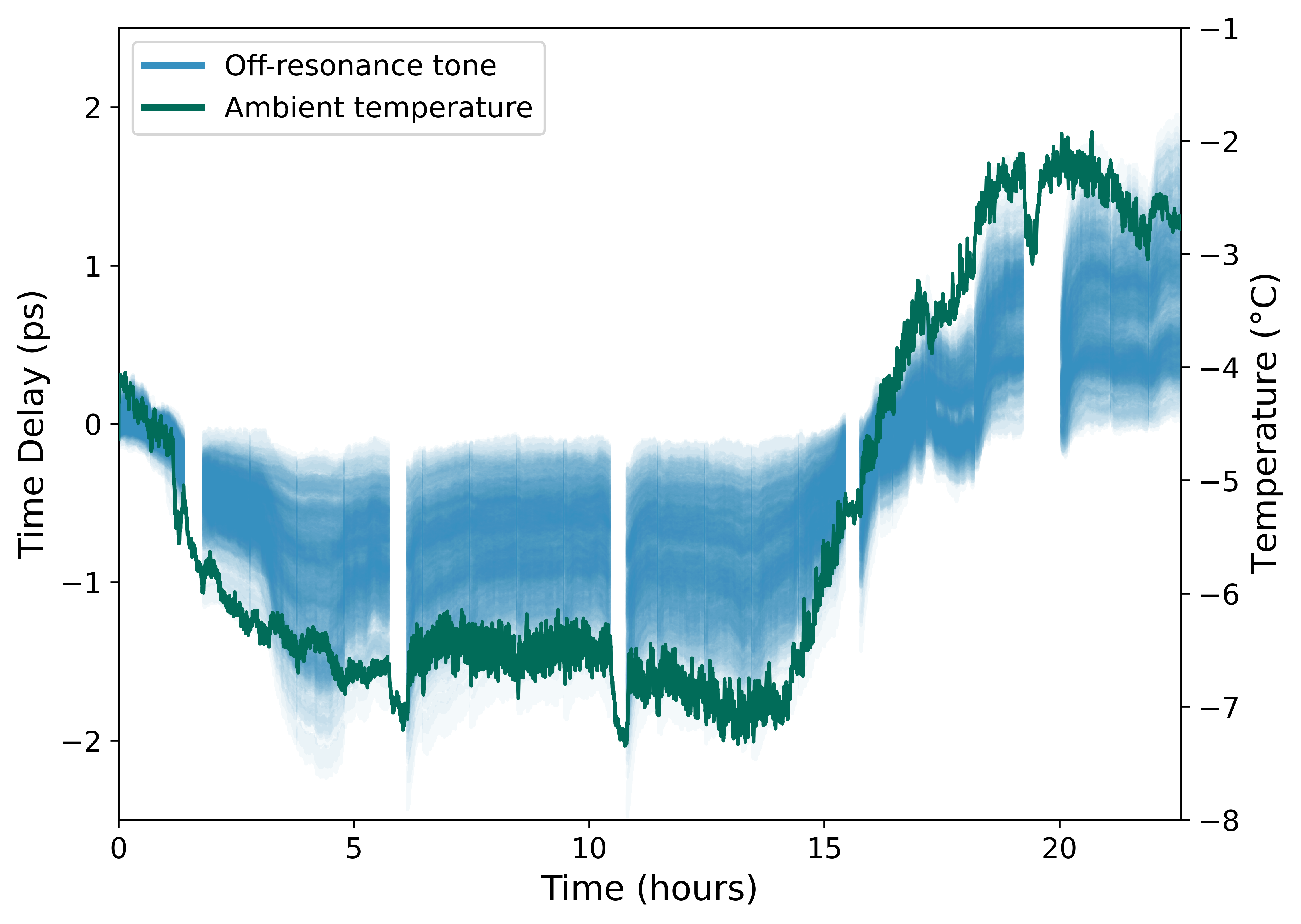}
\caption{Time delay ($\tau$) of off-resonance tones compared with ambient temperature in the LATR cabin during one observing day. Each off-resonance tone is centered to $\tau=0\;\text{ps}$ at the start of the plot.}
\label{fig:tones-temp}
\end{figure}

SMuRF also has the capability to produce ``off-resonance" or ``pilot" tones at locations between the resonant frequencies corresponding to detector channels. These off-resonance tones are not tone-tracked, meaning that between generation and demodulation, they pass through the warm output coaxial cables, the cryogenic circuit, and the warm input coaxial cables unperturbed by interactions with microwave resonators. They can therefore be used to directly measure phase drifts from changes to the warm coaxial cables as effects from within the cryostat are not expected given its thermal stability.

While streaming data from the LATR, we stream data from up to eight off-resonance tones, placed in gaps of at least 10\;MHz between adjacent microwave resonators, per 500\;MHz SMuRF band. This yields up to 64 off-resonance tones per UFM, or 1,152 across the LATR, however in practice we have fewer tones as there are not always eight 10\;MHz gaps in resonant frequencies within each SMuRF band; most resonators are spaced 1-2\;MHz apart. As these tones are not tone-tracked, the output from SMuRF is the angle $\theta=\tan^{-1}(Q/I)$ between the in-phase $\left(I=\text{Re}\left[S_{21}\right]\right)$ and quadrature $\left(Q=\text{Im}\left[S_{21}\right]\right)$ components of the individual off-resonance tones. Following \cite{ToneTracking-SilvaFeaver_2022}, we can use this angle to compute the time delay $\tau$ of the off-resonance tone signal given its frequency $f$:

\begin{equation}
    \tau=\frac{\theta}{2\pi f}
    \label{eqn:phase-delay}
\end{equation}

We also collect data from a thermometer underneath the rear of the LATR which measures the ambient temperature within the receiver cabin a few meters away from the SMuRF electronics and their cables.

\begin{figure}[!t]
\centering
\includegraphics[width=2.75in]{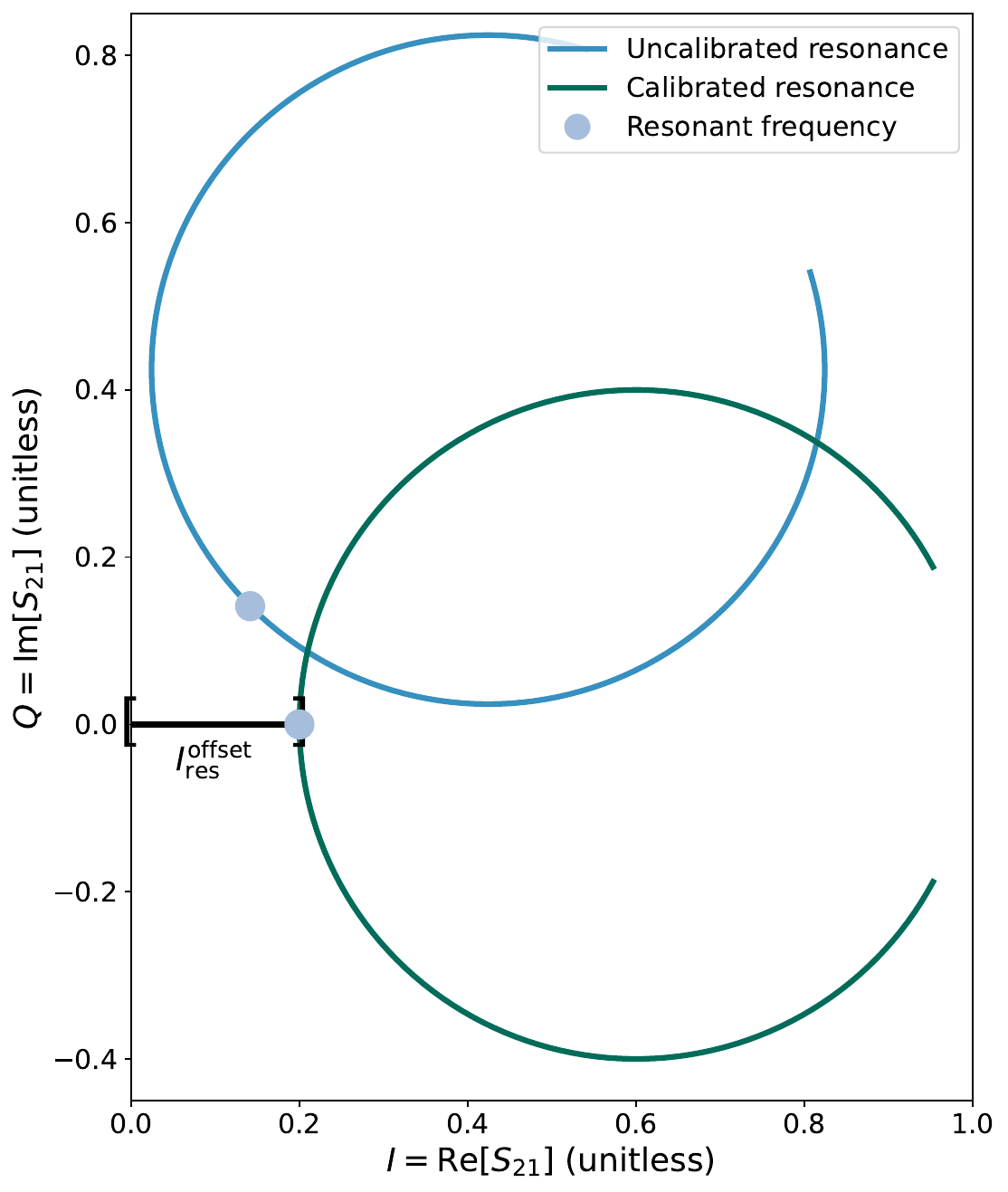}
\caption{In-phase $\left(I=\text{Re}\left[S_{21}\right]\right)$ and quadrature $\left(Q=\text{Im}\left[S_{21}\right]\right)$ components of microwave resonators as tracked by SMuRF before (blue) and after (green) calibration, which corresponds to a rotation in the $IQ$-plane. Resonant frequency locations marked with lilac dots. Distance along the $I$-axis from the resonant frequency to the origin after calibration is labeled as $I_\text{res}^\text{offset}$.}
\label{fig:eta-rotation}
\end{figure}

Figure \ref{fig:tones-temp} shows the measured time delay of 290 of these off-resonance tones generated by 30 AMCs over a period of 22.6 hours of data collection beginning just after sunset on June 2, 2025, over-plotted with the ambient temperature. As each tone begins at a random offset ($\theta$), the tones are centered to $\tau=0\;\text{ps}$ at the start of the plot. Gaps in data collection exist due to routine detector operations occurring alongside changes in telescope pointing in elevation, such as $I-V$ curves. The ambient temperature signal has been filtered to remove anomalous values in excess of $\pm50\degree\;\text{C}$ and passed through a 0.1\;Hz Butterworth low-pass filter.

We observe a correlation between the time delays of the off-resonance tones and the ambient temperature. However, we find that the overall magnitude of this effect is at the level of a few picoseconds over the course of one diurnal swing. We find that the time delays are clustered by the individual RF transmission line circuits, and that there is spread within each circuit due to the 2\;GHz range in off-resonance tone frequency.

\begin{figure}[!t]
\centering
\includegraphics[width=3.25in]{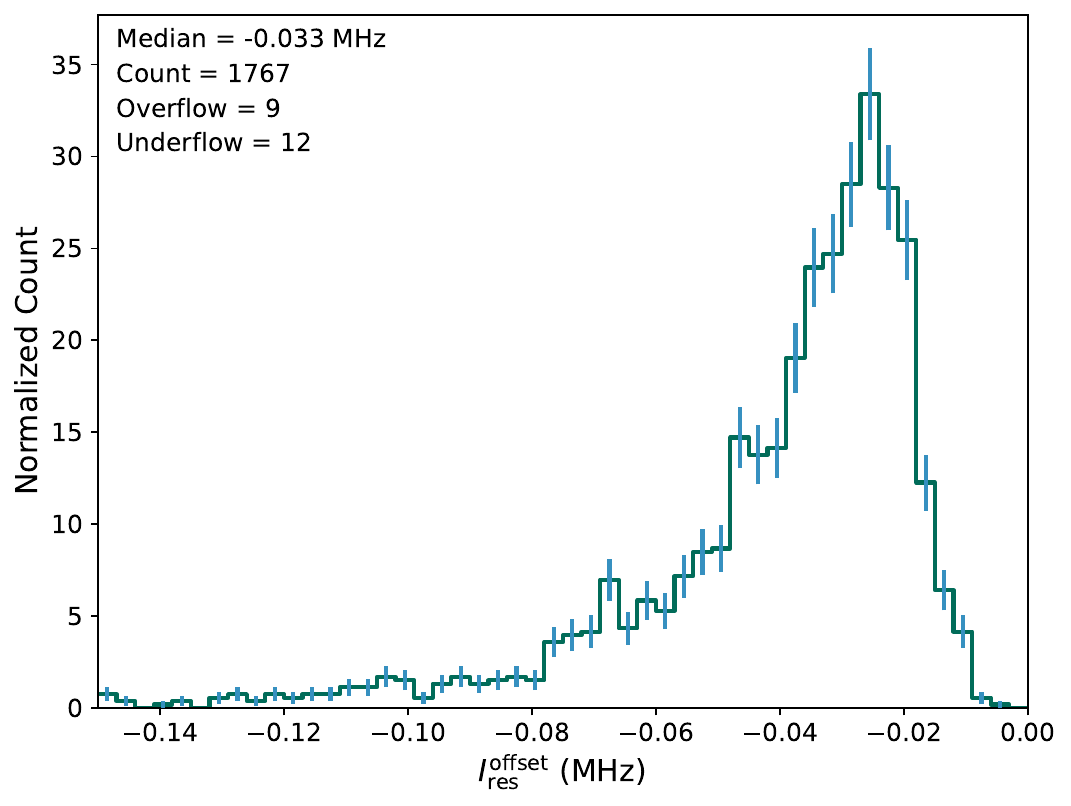}
\caption{Distribution of $I_\text{res}^\text{offset}$, as defined in Figure \ref{fig:eta-rotation}. Measured using SMuRF for microwave resonators deployed to one UFM installed to the LATR. Error bars show $1\sigma$ Poisson counting uncertainties.}
\label{fig:dI}
\end{figure}

To analyze whether this effect could lead to a CMB signal contamination, we compute the inferred detector noise-equivalent current from the phase drift signal. Microwave resonators trace curves in the $IQ$-plane as shown in Figure \ref{fig:eta-rotation}. SMuRF uses a firmware-level calibration to rotate these curves such that perturbations about the resonant frequency are measured as changes entirely in $Q$, as shown by the green curve. For a more in-depth discussion of this calibration, refer to \cite{SMuRF-Yu_2023}. The RF phase drift causes small rotations of this circle by angle $\theta$, which lead to shifts in $Q$ that mimic a changing TES detector signal as it would be read out via the flux-ramped SQUIDs. In the limit of small $\theta$, the signal contaminant $\delta Q$ is:

\begin{equation}
    \delta Q=I_\text{res}^\text{offset}\tan\theta
    \label{eqn:phase-rotation}
\end{equation}

Where $I_\text{res}^\text{offset}$ is the post-calibration distance along the $I$-axis from the resonant frequency to the origin, as labeled in Figure \ref{fig:eta-rotation}. Figure \ref{fig:dI} shows the distribution of $I_\text{res}^\text{offset}$ as measured for the microwave resonators deployed to one of the LATR's UFMs, following a calibration of $Q$ to units of resonator frequency offset. We observe a median value of $-3.31\times10^4\;\text{Hz}$.

To refer this change in $Q$ to a corresponding change in current through the TES detectors $I_\text{TES}$, we analyze measured SQUID curves. This measurement is discussed in more detail in the Appendix. For SQUIDs installed to one UFM containing $90\;\text{GHz}$ and $150\;\text{GHz}$ detectors deployed to the LATR, we measure a median value for $\langle df/dI_\text{TES}\rangle_{\Phi_0}$ of $5.71
\times10^{-2}\;\text{Hz\;pA}^{-1}$.

\begin{figure}[!t]
\centering
\includegraphics[width=3.25in]{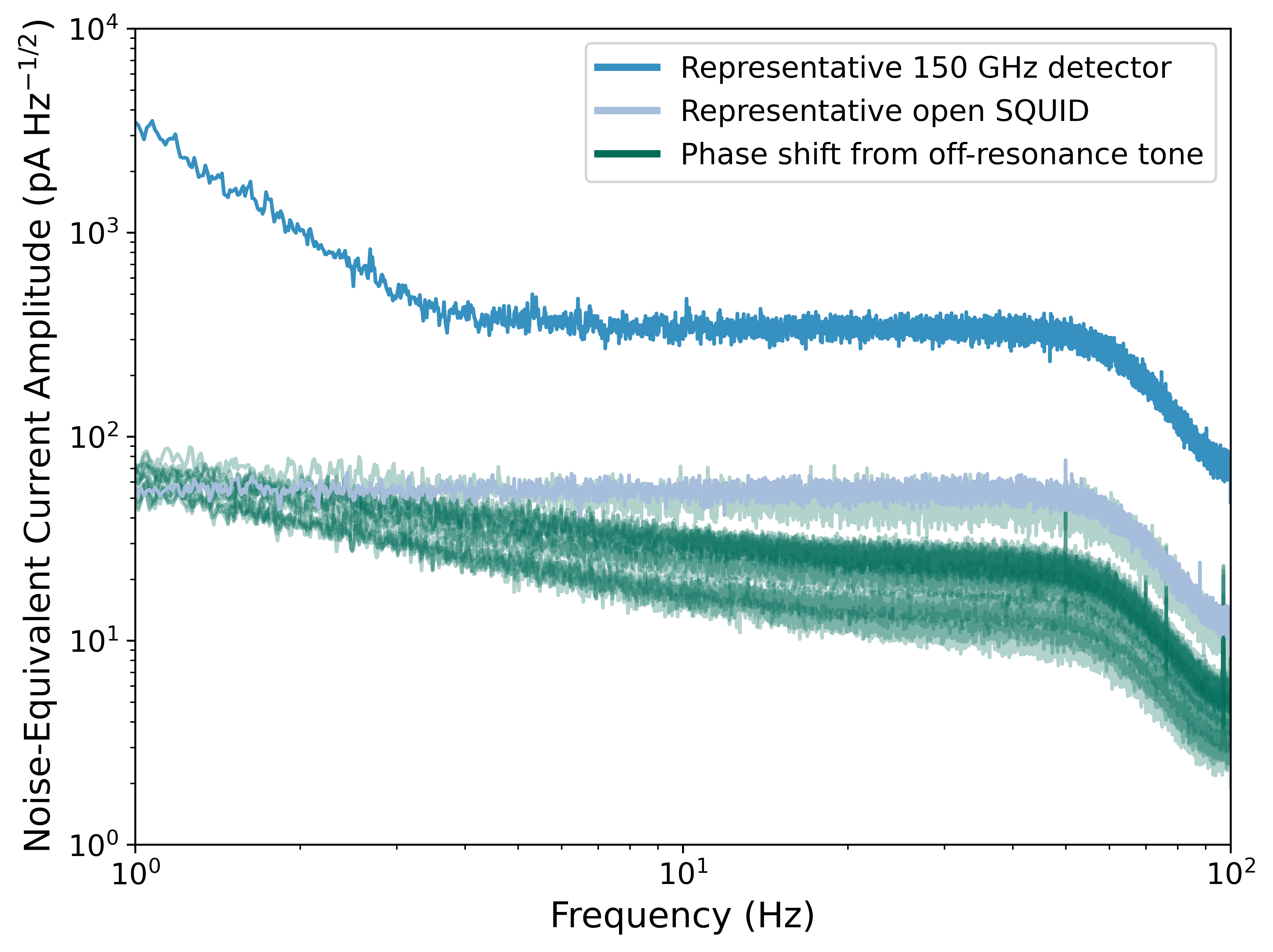}
\caption{Amplitude spectral densities of the inferred TES noise-equivalent currents from the measured phase drifts (green) during a single 40-minute-long constant-elevation scan with one UFM installed to the LAT. Shown in comparison to a representative 150\;GHz on-sky in-transition detector (blue) and a representative open SQUID channel (lilac).}
\label{fig:tones-asd}
\end{figure}

Using this conversion, we analytically compute a distribution of the noise-equivalent current of this phase drift effect. For a more accurate conversion, a future study would simulate the effect of flux ramp demodulation, such as by using the \texttt{babysmurf} software package\cite{BabySMURF-Yu_2022}. Figure \ref{fig:tones-asd} shows the amplitude spectral densities (ASDs) of the off-resonance tones used to probe one UFM installed to the LAT during a single 40-minute-long constant elevation scan at $40\degree$, converted using the aforementioned median measured values of $I_\text{res}^\text{offset}$ and $\langle df/dI_\text{TES}\rangle_{\Phi_0}$ for that UFM.

We compare these ASDs (green) to that of a representative 150\;GHz on-sky in-transition detector (blue) and of a microwave resonator which is coupled to a SQUID which is not itself coupled to a TES detector, known as an ``open SQUID" channel (lilac). The open SQUID channel provides a probe of readout noise as there are no contributions to its signal from a detector. We observe that the calculated noise contribution from the phase drift effect is not a significant driver of the overall on-sky detector noise, and that it is within the readout noise budget. As the readout noise levels of the LATR's detectors have already been shown to be largely within specification, we do not expect this phase drift from the warm RF cables to affect the telescope's mapping speed\cite{Satterthwaite-SMuRFDeployment}.

\section{Conclusion}
\label{sec:conclusion}
\noindent

The room-temperature components of the SMuRF systems which enable the warm readout of the Simons Observatory's LAT's nearly 31,000 detectors are susceptible to diurnal temperature changes. Though this effect induces a systematic phase drift in the RF transmission lines which is correlated with the ambient temperature, the magnitude of this phase drift effect is at the picosecond level over the course of one day-night cycle. Using an analytical conversion, we find that this effect is a small contribution to the on-sky detector noise of the LATR, the readout-derived component of which has already been shown to be within the specification for successful operation of the telescope.

\appendix{}
\label{sup:squid-curves}
\noindent

\begin{figure}[!t]
\centering
\includegraphics[width=3.25in]{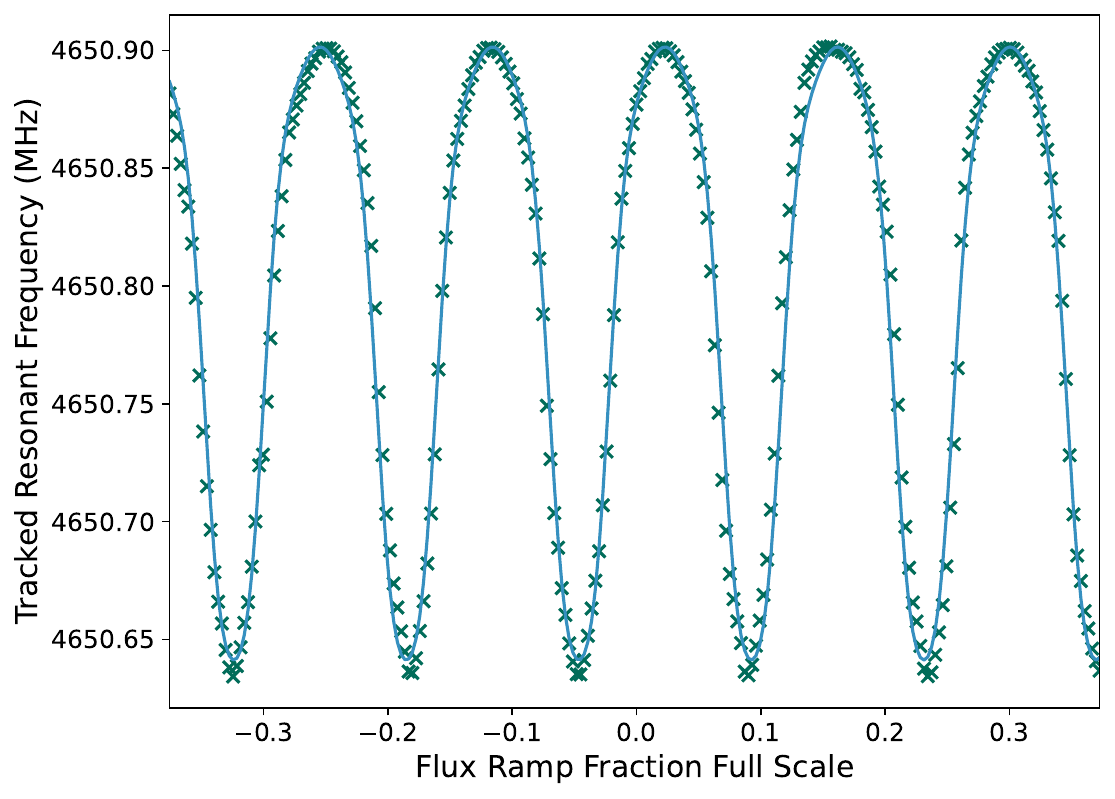}
\caption{Sample SQUID curve with measured points shown as green marks and the fit curve, using Equation \ref{eqn:squid-curve}, drawn in blue.}
\label{fig:squid-curve}
\end{figure}

\begin{figure}[!t]
\centering
\includegraphics[width=3.25in]{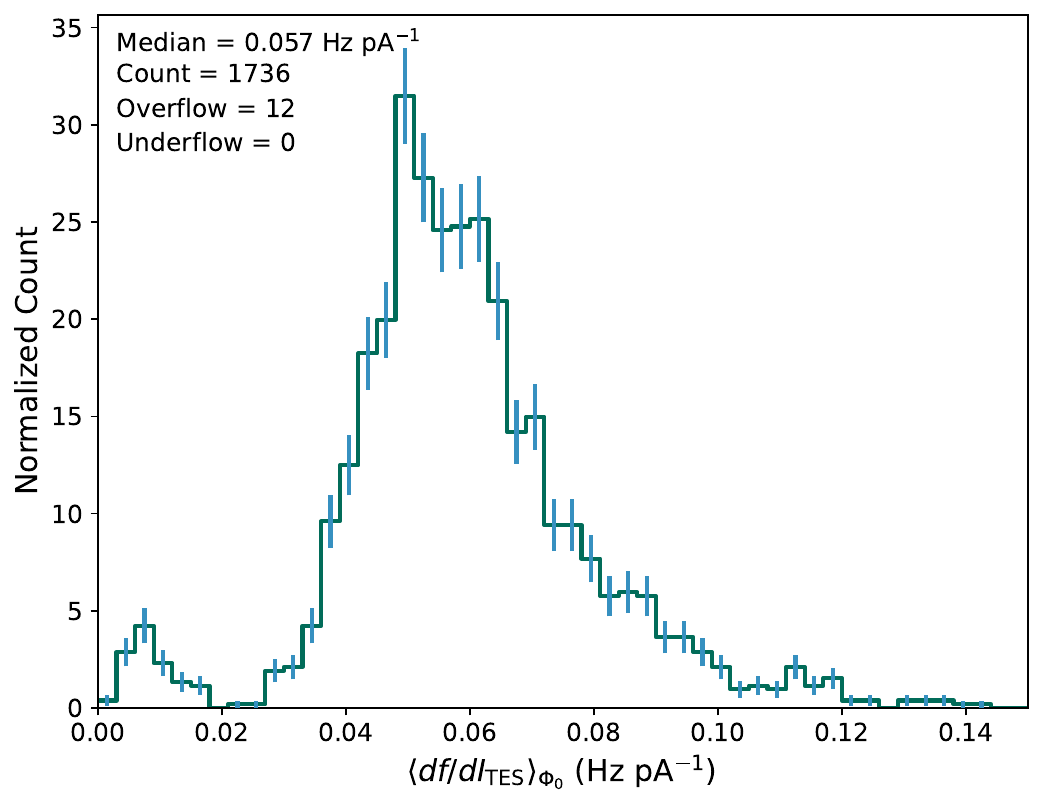}
\caption{Distribution of $\langle df/dI_\text{TES}\rangle_{\Phi_0}$. Measured using SMuRF for SQUIDs in one 90\;GHz and 150\;GHz UFM installed to the LATR. Error bars show $1\sigma$ Poisson counting uncertainties.}
\label{fig:dfdI}
\end{figure}

In order to infer the change in TES current associated with a change in frequency of the flux-ramped signal read out through the SQUID on the UFM multiplexing circuit, we measure the response of this frequency to varying flux ramp signals. This traces a ``SQUID curve" whose properties we analyze.

Following \cite{Thesis-SilvaFeaver_2023}, we fit this curve to the sum of five cosine harmonics:

\begin{equation}
    f(\phi)=\alpha+\frac{1}{2}\sum_{i=1}^5\beta_i\cos\left(2\pi i(\phi-\phi_0)/\sigma\right)
    \label{eqn:squid-curve}
\end{equation}

Where $f$ is the frequency of the microwave resonator as a function of flux through the SQUID $\phi$. The model is parameterized by a constant $\alpha$, amplitude $\beta_i$ of harmonic $i\in[1,5]$, offset $\phi_0$, and scale $\sigma$.

Figure \ref{fig:squid-curve} shows a sample SQUID curve for one detector channel as its flux ramp is varied between $\pm0.3755$ of its full scale. From this, we numerically compute the gradient $df/d\phi$ and average over the course of one SQUID flux quantum $\left(\Phi_0\right)$, which induces a full swing in resonator frequency, in order to estimate $\langle df/d\phi\rangle_{\Phi_0}$.

We convert the change in SQUID flux ($d\phi$) to an associated change in current through the TES ($dI_\text{TES}$) using the mutual inductance between the SQUID and the TES ($227\;\text{pH}$) and the magnetic flux quantum ($2.068\times10^{-15}\;\text{Wb}$)\cite{SQUID-Dober_2021}. Figure \ref{fig:dfdI} shows the distribution of $\langle df/dI_\text{TES}\rangle_{\Phi_0}$ estimated using these SQUID curves measured on a single 90\;GHz and 150\;GHz UFM installed to the LATR. Curves whose derived peak-to-peak swing in frequency is less than 10\;kHz, signifying a poor fit to the model, are omitted. We observe a median value of $5.71\times10^{-2}\;\text{Hz\;pA}^{-1}$.

\section*{Acknowledgments}
\noindent
This work was supported in part by the Department of Energy, Laboratory Directed Research and Development program at SLAC National Accelerator Laboratory, under contract DE-AC02-76SF00515. This work was also supported in part by a grant from the Simons Foundation (Award \#457687, B.K.). This work was also supported by the U.S. National Science Foundation (Award Number: 2153201). Several figures in this paper were created using the Python packages \texttt{numpy} and \texttt{matplotlib} \cite{numpy,matplotlib}.

\bibliographystyle{IEEEtran}
\bibliography{references}

\end{document}